  \documentstyle[12pt]{article}
  
\catcode`@=11
\def\secteqno{\@addtoreset{equation}{section}%
\def\theequation{\thesection.\arabic{equation}}}
\catcode`@=12
\secteqno
\newcommand{\be}{\begin{equation}}
\newcommand{\ee}{\end{equation}}
\newcommand{\bea}{\begin{eqnarray}}
\newcommand{\eea}{\end{eqnarray}}
\newcommand{\bref}[1]{(\ref{#1})}
\newcommand{\ep}{\epsilon} 
\newcommand{\T}{\theta} 
   
\newcommand{\A}{\alpha} \newcommand{\B}{\beta}
\newcommand{\G}{\Gamma} \newcommand{\D}{\delta}

           \newcommand{\s}{\sigma}
          
\newcommand{\z}{\zeta}
           \newcommand{\Th}{\Theta}

\def\pa{\partial}

\newcommand{\nn}{\nonumber}

\def\CL{{\cal L}}

\def\CF{{\cal F}}
\def\Tb{{\overline\theta}}

\def\tension{{{\rm T}_{(p)}}}
\def\tensionm{{{\rm T}}}

\def\ba{\overline}

\def\tp{\tilde p}
\def\t{\tilde}

\def\l{{\ell}}
\def\gg{{\Gamma_{11}}}

\newcommand{\slcP}{/ {\hskip-0.27cm{\cal P}}}

\newcommand{\sldX}{/ {\hskip-0.27cm{dX}}}
\newcommand{\sltp}{/ {\hskip-0.27cm{\tilde{p}}}}
\newcommand{\slPi}{/ {\hskip-0.27cm{\Pi}}}

\newcommand{\slV}{/ {\hskip-0.21cm{V}}}

\newcommand{\slVt}{/ {\hskip-0.21cm{\tilde V}}}
\newcommand{\CC}{{c}}
\begin{document}
\vfill
\vbox{
\hfill  April 13, 1998 \null\par
\hfill TOHO-FP-9759\null\par
\hfill }\null
\vskip 20mm
\begin{center}
{\bf \Large Wess-Zumino actions for IIA D-branes}\par
{\bf \Large and their supersymmetries}\par
\vskip 10mm
{\large Machiko\ Hatsuda and Kiyoshi\ Kamimura$^\dagger$}\par
\medskip
{\it Department of Radiological Sciences,
Ibaraki Prefectural University of Health Sciences,\\
\ Ami\ Inashiki-gun\ Ibaraki\ 300-0394, Japan \\
$^\dagger$ Department of Physics, Toho University,
\ Funabashi\ 274-8510, Japan
}\par
\medskip
\vskip 10mm
\end{center}
\vskip 10mm
\begin{abstract}
We present Wess-Zumino actions for general IIA D-p-branes 
in explicit forms. 
We perform the covariant and irreducible separation of the fermionic 
constraints of IIA D-p-branes 
into the first class and the second class.
A necessary condition which guarantees this separation is discussed.
The generators of the local supersymmetry (kappa symmetry) and the 
kappa algebra are obtained. 
We also explicitly calculate the conserved charge of the global supersymmetry
 (SUSY) and the SUSY algebra which contains topological charges.  
\end{abstract}
\par

\newpage
\setcounter{page}{1}

\parskip=7pt

\section{ Introduction}\par
\indent

D-brane dynamics play an important role in the non-perturbative 
superstring physics.
The characteristic feature of the D-branes is their non-zero Ramond/Ramond 
(RR) charges \cite{pol}.
The RR coupling is realized through the Wess-Zumino action 
\cite{Doug,BRoo,BT,GHT}.
The D-branes are representation of
 the global supersymmetry (SUSY) algebra with central extension
whose origin is the Wess-Zumino action \cite{AGIT}.
The Wess-Zumino action is essential for 
the D-brane dynamics showing rich structures of various dualities, 
and its explicit expression will be important especially for the D-brane 
quantization.
Wess-Zumino actions of the superstring theories are required
from the kappa invariance \cite{GS}. 
The actions for D-p-branes with the kappa symmetry have been proposed 
in references \cite{Shgf,Cw,BT}.
In these references a differential equation of the Wess-Zumino action
is obtained from the requirement of the kappa symmetry,
and explicit expressions of the Wess-Zumino actions are presented only 
for small p cases. 
In order to figure out a suitable treatment of fermionic variables,
one needs an explicit expression of the fermionic constrains.
In a previous paper we have given a concrete expression of the Wess-Zumino 
action by solving the differential equation for general 
IIB D-p-branes \cite{KH}.
It has a compact and closed form and enables us to confirm
algebraic properties of the local and the global symmetries.
In this paper we have completed this program for IIA D-p-branes.

In section 2, we derive the Wess-Zumino action for general IIA D-p-branes 
in an explicit form.
In section 3, the IIA D-p-brane actions are analyzed in canonical formalism. 
Covariant and irreducible separation of the fermionic constraints into the first class and the second class can be performed. 
We examine necessary conditions which guarantee this separation.
The generators of the local supersymmetry (kappa symmetry)
 and the kappa algebra are obtained.
In section 4, we calculate explicitly the conserved charge of 
the global supersymmetry (SUSY) and the SUSY algebra which contains 
topological charges. 
The physical interpretations of topological charges are discussed 
in the last section.

\vskip 6mm

\section{Wess-Zumino action for type IIA D-branes}
\indent

A Wess-Zumino action for a D-p-brane is obtained
as a $p+1$ form part of a symbolic sum  of differential forms 
\bea
L^{WZ}~&=&~~\tensionm~C~e^{\cal F},
\label{WZ}
\eea
where  $C$ is the RR gauge field and $\tensionm$ is the D-brane tension 
\cite{Doug,BRoo,BT,GHT}.
$\CF$ is the supersymmetric 
extension of the Born-Infeld U(1) field strength, 
\bea
\CF~=~dA~-~B~
\eea
where $B$ is the~NS-NS(NS)~ 2-form. 
Denoting $H$ as the field strength of $B$,
$~ H=dB$,
 $R$ is the field strength of RR potential $C$,~
\bea
R~=~dC~-~H~C ~~.
\label{curvature}
\eea
It satisfies the Bianchi identity,
\be~dR-HR=0\quad.
\label{Bianchi}
\ee
Then the  Wess-Zumino action satisfies 
\bea
d~L^{WZ}~&=&~~\tensionm~R~e^{\cal F}~~.
\label{dWZ}
\eea
The form of the curvature $R$ is determined by the requirement
of kappa invariance of the total action. 
On the other hand the form of the RR potential $C$ is not unique 
but determined up to 
RR gauge transformations
\bea
C~~\rightarrow~~C'~=~C~+~d~\Lambda~-~H~\Lambda.
\label{RRgaugetrans}
\eea
Under the RR transformation, $R$ is invariant while
the Wess-Zumino action changes by an exact form,
\bea
L^{WZ}~~\rightarrow~~{L^{WZ}}'~=~L^{WZ}~+~d~[~\Lambda~e^\CF~].
\eea

The Wess-Zumino action is obtained
by solving the differential equation \bref{dWZ},
or equivalently by solving \bref{curvature} for the RR potential $C$.
The type IIA super D-p-brane is 
described by worldvolume fields; 
the 10-dimensional coordinates $X^m$,
32 components Majorana fermion $\T$ and the U(1) gauge field $A_\mu$.
The kappa invariance requires the form of the equation \bref{dWZ}
\bea
d~L^{WZ}~&=&~~\tensionm~R~e^{\cal F},~~~~~~~~~~~~
R~=~d\Tb~{\cal C}_A~d\T,
\label{dWZIIA}
\eea
where  
$~{\cal C}_A~$ is introduced in the ref.\cite{Shgf}
together with $~{\cal S}_A~$,
\bea
{\cal C}_A(\slPi)&=&\sum_{\l=0}~
(\Gamma_{11})^{\l+1}~~\frac{\slPi^{2\l}}{(2\l)!}~~~~=~
\Gamma_{11}~+~\frac{\slPi^2}{2!}~+~\Gamma_{11}~\frac{\slPi^4}{4!}~+~...~,
\\
{\cal S}_A(\slPi)&=&\sum_{\l=0}~
(\Gamma_{11})^{\l+1}\frac{\slPi^{2\l+1}}{(2\l+1)!}~=~
\Gamma_{11}\slPi~+~\frac{\slPi^3}{3!}~+~\Gamma_{11}\frac{\slPi^5}{5!}~+~...~.
\label{defS}
\eea
$R$ is manifestly invariant under the SUSY transformation.
The NS two form $B$ and its field strength $H$ are
\bea
B&=&-~\Tb~\gg~\Gamma\cdot (\Pi-\frac12\Tb\G d\T)~d\T ,
\\
H&=&dB~=~-~d\Tb~\gg~\Gamma\cdot \Pi~d\T.
\eea
\medskip
In this section we will integrate \bref{dWZIIA}
 to get the Wess-Zumino action, 
or equivalently the RR potential $C$, explicitly
in an analogous manner to the type IIB case \cite{KH}.
Several formulas of type IIA case are obtained from those of type IIB 
ones by replacing $\tau_3$ by $\gg$. 
However there is a difference that $\tau_3$ commutes with $\Gamma$
matrices while  $\gg$  anti-commutes with $\Gamma$'s.
It makes the IIA formulas more involved.

In order to recognize the cyclic identity easier, 
it is convenient to introduce  $\Gamma$ matrix valued 10D vector for any
type IIA spinors $\psi$ and $\phi$
\bea
(V^m_{\psi,\phi})^\alpha _\beta&\equiv&
\delta^\alpha _\beta(\ba\psi~\G^m~\phi)~
+~(\Gamma_{11})^\alpha _\beta~(\ba\psi~\Gamma_{11}~\G^m~\phi).
\label{BBB}
\eea
We define "slashed" quantities by contracting with $\Gamma$ from the right
\bea
\slV_{\psi,\phi}~\equiv~V_{\psi,\phi}^m~\Gamma_m~
\eea
$V^m$ contracted with $\Gamma_m$ from the left defines $\t V^m$ as
\bea
\t\slV_{\psi,\phi}~\equiv~\Gamma_m~V_{\psi,\phi}^m~=~
\t V_{\psi,\phi}^m~\Gamma_m~\equiv~
\{(\ba\psi~\G^m~\phi)~-~
\Gamma_{11}(\ba\psi~\Gamma_{11}~\G^m~\phi)\}~~\Gamma_m.
\eea
Using this notation the cyclic identity holds as
\bea
\slV_{\psi,\phi}~\T~
+~\slV_{\phi,\T}~\psi~+~\slV_{\T,\psi}~\phi~=~0
\label{cyc}
\eea
for three odd IIA spinors. It frequently appears an expression
\bea
V^m&\equiv&
V^m_{\T,d\T}.
\eea
The cyclic identity \bref{cyc} tells, for example, 
\be
d\slV~\T~+~2~\slV~d\T~=~0~~~~\rightarrow~~~~~
d\slV~d\T~=~0,~~~~~V~\cdot~dV~=~0.
\label{slbdt}
\ee

The well-known IIA identities are rewritten in terms of $\slV$ variables.
For example, 
\bea
&&\sum_{{\rm perm } \psi,\phi,\theta,\chi}~~
\left[ (\bar{\psi}\Gamma^m \phi)~(\bar{\theta}\Gamma_{mn}\chi)
+(\bar{\psi}\Gamma_{11} \phi)~(\bar{\theta}\Gamma_{11}\Gamma_{n}\chi)
\right]\nn\\
&&~~~~~~=
\sum_{{\rm perm } \psi,\phi,\theta,\chi}~~
\left[ (\bar{\psi}_1\Gamma^m \phi_1)~
(\bar{\theta}_1\Gamma_{mn}\chi_2+\bar{\theta}_2\Gamma_{mn}\chi_1
+\eta_{mn}\bar{\theta}_1\chi_2-\eta_{mn}\bar{\theta}_2\chi_1)
+~1\leftrightarrow 2~
\right]\nn\\
&&~~~~~~=
\sum_{{\rm perm }  \psi,\phi,\theta,\chi}~~
(\bar{\theta}\slV_{\psi,\phi})~\Gamma_n 
\chi~~
=0\quad,\label{IIAcyc}
\eea  
for type IIA spinors $\psi,\phi,\theta,\chi$. 
In the last line of \bref{IIAcyc} 
permutation makes the bracket to vanish 
by the cyclic identity \bref{cyc}.
More general IIA identities \cite{Hammer} are also rewritten
using with the relation
\bea
\Gamma_m\Gamma_{m_1\cdot\cdot\cdot m_q}~=~\Gamma_{mm_1\cdot\cdot\cdot m_q}
+q\eta_{m[m_1}\Gamma_{m_2\cdot\cdot\cdot m_q]}\quad,
\quad {\cal O}_{[m_1\cdot\cdot\cdot m_q]}
=\frac{1}{q!}\sum_{{\rm perm} m's} {\cal O}_{m_1\cdot\cdot\cdot m_q}\quad  ,
\eea 
as
\bea
&&\sum_{{\rm perm } \psi,\phi,\theta,\chi}~~
\left[ 
(\bar{\psi}\Gamma^m \phi)~(\bar{\theta}\Gamma_{mn_1\cdot\cdot\cdot n_{2q-1}}
(\Gamma_{11})^{q-1}\chi)
+(2q-1)(\bar{\psi}\Gamma_{11} \Gamma_{[n_1}\phi)
~(\bar{\theta}\Gamma_{n_2\cdot\cdot\cdot n_{2q-1}]}(\Gamma_{11})^{q}\chi)
\right] \nn\\
&&~~~~~~=
(-1)^{q-1}\sum_{{\rm perm }  \psi,\phi,\theta,\chi}~~
(\bar{\theta}\slV_{\psi,\phi})~\Gamma_{n_1
\cdot\cdot\cdot n_{2q-1}} 
\chi~~
=0\quad\label{IIAidn},\quad q>0
\quad .
\eea  
It is useful to have the following relations
\bea
d~({\cal S}_A~e^{\cal F})&=&\frac12~(~d\slV~\tilde{\cal C}_{A}~+~
{\cal C}_A~d\slVt~)~e^{\cal F},
\nonumber\\
d~({\cal C}_A~e^{\cal F})&=&\frac12~(~-d\slV~\Gamma_{11}~
\tilde{\cal S}_{A}~
+~{\cal S}_A~\Gamma_{11}~d\slV~)~e^{\cal F}~,
\label{dCS}
\eea
where $\tilde{\cal S}_{A}$ and $\tilde{\cal C}_{A}$ are introduced by 
replacing $\Gamma_{11} ~\rightarrow~(-\Gamma_{11})$ in
${\cal S}_A$ and ${\cal C}_A$ of \bref{defS}.

\medskip

Next we define j-form IIA spinor $ {\Th}_j$ by
\bea
 {\Th}_{j}&\equiv& \hat{\slV}_j~ {\Th}_{j-1},~~~~(j\geq 1),~~~~~~~
 {\Th}_0~\equiv~ \T,
\eea
where 
\bea
\hat{V}_j^m=
\left\{
\begin{array}{ll}
V^m & {\rm for\  odd } ~j,\\
\tilde{V}^m & {\rm for\ even }~  j.
\end{array}\right.~~
\eea
The concrete forms are
\bea
{\Th}_0&=&~ \T \nonumber\\
 {\Th}_1&=&~\slV~\T \nonumber\\
 {\Th}_2&=&~\slVt~\slV~\T 
\nonumber\\
 {\Th}_3&=&~\slV~\slVt~\slV~\T\nonumber\\
\cdot\cdot\cdot
\eea
It has the $j+1$ parity and it holds a useful relation,
\be
d~ {\Th}_j~=~~\frac{2j+1}{2}~d\hat\slV~ {\Th}_{j-1}~,~~~~~~~~~(j=1,2,...).
\label{dTh}
\ee
The important quantities describing the D-branes 
are odd IIA spinor form $ \Th_S$ and even one $ \Th_C$ defined by
\bea
 {\Th}_S&\equiv&-\sum_{n=0}~\frac{1}{(4n+3)!!}~(-\Gamma_{11})^{n+1}~
 {\Th}_{2n+1}~=~
\frac{1}{3!!}~\Gamma_{11}~ {\Th}_1~-~\frac{1}{7!!}~ {\Th}_3~+...,
\nn\\
 {\Th}_C&\equiv&\sum_{n=0}~\frac{1}{(4n+1)!!}~~(\Gamma_{11})^{n}~
 {\Th}_{2n}~~~=~
\T~+~\frac{1}{5!!}~\Gamma_{11}~ {\Th}_2~+~... ~.
\label{defThCS}
\eea
They satisfy, using \bref{dTh}
\bea
d  {\Th}_S&=&\frac{1}{2}\Gamma_{11}d\slV\Theta_C ,\nn
\\
d  {\Th}_C&=&d\T~-~\frac{1}{2}d\tilde{\slV}\Theta_S\quad .
\label{dThCS}
\eea
\medskip

Now we are ready to integrate \bref{dWZIIA}
to find the $L^{WZ}$.
We first write the right hand side of \bref{dWZIIA} as
\bea
&&d\Tb~{\cal C}_A~e^{\cal F}~d\T=d~L_1~+~I_1,~~~~~~
L_1~=~d\Tb~{\cal C}_A~e^{\cal F}~\T,~~~~~
I_1~=~-d\Tb~d~({\cal C}_A~e^{\cal F})~\T~.\nonumber\\
&&
\eea
Using \bref{dCS}, \bref{slbdt} and \bref{dTh}
\bea
I_1&=&-\frac12 d\Tb~{\cal S}_A~e^{\cal F}~\Gamma_{11}~ d\slV~\T
=~-\frac13~d\Tb~{\cal S}_A~e^{\cal F}~\Gamma_{11}~d {\Th}_1~
\equiv~d~L_2~+~I_2\nonumber\\
&&L_2~=~\frac13~d\Tb~{\cal S}_A~e^{\cal F}~\Gamma_{11}~ {\Th}_1~,~~~~~
I_2~=~-\frac13~d\Tb~d({\cal S}_A~e^{\cal F})~\Gamma_{11}~ {\Th}_1~.
\eea
Repeating this procedure  ( actually it terminates
at $(p+1)$-th step for the $p$ brane ) and summing up $L_j$'s, 
we arrive at a compact expression for the Wess-Zumino action:
\bea
L^{WZ}&=&\tensionm~
d\Tb(~{\cal C}_A~ {\Th}_C+~{\cal S}_A~ {\Th}_S)e^{\cal F}.
\label{solwz}
\eea
The Wess-Zumino actions of the D-p-brane is the $(p+1)$ form part of 
\bref{solwz}.
It is easy to check that \bref{solwz} satisfies \bref{dWZIIA}
by using the relations \bref{dCS} and \bref{dThCS} with \bref{slbdt}.

\medskip
Corresponding to the fact that the lagrangian is determined up to
total divergence, the RR potential $C$ is obtained from \bref{solwz} 
up to RR gauge transformations \bref{RRgaugetrans}.  
For a D-2-brane the Wess-Zumino action given from  \bref{solwz} is
\bea
{\cal L}_{3}^{WZ}=T_{(2)}~[~
 \frac{1}{2} d\bar{\theta}~\slPi^2~\theta~
-~\frac{1}{3} ~d\bar{\theta}~\slPi~ {\Theta}_1~
+~\frac{1}{15}~ d\bar{\theta}~ {\Theta}_2~
+~d\bar{\theta}~\Gamma_{11}~\theta~{\cal F}~]. 
\label{kyswz2}
\eea
It differs from, for example, one in the ref.
\cite{Hammer} 
by a RR transformation \bref{RRgaugetrans} with $\Lambda$
\bea
\Lambda~=~
\frac{1}{90}\bar{\theta} {\Theta}_2
+\frac{1}{12}(\Pi^m-\frac13\bar{d\theta}\Gamma^m\theta)
\bar{\theta}\Gamma_m {\Theta}_1~.
\eea
It is noted that $\Lambda$ 
does not have explicit dependence on $X$ and $A$ but on $\Pi$ and $dA$. 
\medskip

We summarize our results by adding the previous results for type IIB 
\cite{KH} in Table 1.

\begin{table}[hbtp]
\caption[RC]{Summary of type IIA and IIB D-branes}
\label{RRC}
\begin{center}
\begin{tabular}{|c|c|c|}
\hline
 &IIA&IIB\\
\hline
&&\\
$B$~&$~-~\Tb~\gg~\Gamma\cdot (\Pi-\frac12\Tb\G d\T)~d\T
$&$~-~\Tb~\tau_3~\Gamma\cdot (\Pi-\frac12\Tb\G d\T)~d\T
$\\
\hline
&&\\
$H$~&$~-~d\Tb~\gg~\Gamma\cdot \Pi~d\T
$&$-~d\Tb~\tau_3~\Gamma\cdot \Pi~d\T
$\\
\hline
&&\\
$C$~&$~d\Tb(~{\cal C}_A~ {\Th}_C+~{\cal S}_A~ {\Th}_S)~$&
$~d\bar{\theta}~\tau_1(~{\cal S}_B~\Th_C+{\cal C}_B~\Th_S~)~$\\
\hline
&&\\
$R$~&$~d\Tb~{\cal C}_A~d\T~$&$~d\Tb~{\cal S}_B~\tau_1~d\T~$\\
\hline
&&\\
${\cal C}_{A,B}$~&
$~\sum_{\l=0}~
(\Gamma_{11})^{\l+1}~\frac{\slPi^{2\l}}{(2\l)!}$&
$\sum_{\l=0}~
(\tau_{3})^{\l+1}~\frac{\slPi^{2\l}}{(2\l)!}$\\
\hline
&&\\
${\cal S}_{A,B}$~&
$~\gg~\sum_{\l=0}~
(\Gamma_{11})^{\l}\frac{\slPi^{2\l+1}}{(2\l+1)!}
$&
$~\sum_{\l=0}~
(\tau_{3})^{\l}\frac{\slPi^{2\l+1}}{(2\l+1)!}
$\\
\hline
&&\\
${\Th}_{C}$~&
$\sum_{n=0}~\frac{1}{(4n+1)!!}~(\Gamma_{11})^{n}~ {\Th}_{2n}
$&$
\sum_{n=0}~\frac{1}{(4n+1)!!}~(-\tau_{3})^{n}~{\Th}_{2n}
$\\
\hline
&&\\
${\Th}_{S}$~&
$-\sum_{n=0}~\frac{1}{(4n+3)!!}~(-\Gamma_{11})^{n+1}~ {\Th}_{2n+1}
$&$
\sum_{n=0}~\frac{1}{(4n+3)!!}~(-\tau_{3})^{n+1}~{\Th}_{2n+1}
$\\
\hline
\end{tabular}
\end{center}
\end{table}

\eject

\section{Local supersymmetry (kappa symmetry)}
\indent

In this section we derive constraint equations in the canonical formalism,
and examine the correct treatment of the fermionic constraints.
The lagrangian density of the system is 
\be
\CL~=~\CL^{DBI}~+~\CL^{WZ},
\ee
\bea
\CL^{DBI}&=&-~\tension~\sqrt{-\det(G_{\mu\nu}~+~\CF_{\mu\nu})},~
\nn\\
\CL^{WZ}&=&~[L^{WZ}]_{p+1},~~~~~~~
L^{WZ}=\tension~d\bar{\theta}~
({\cal C}_A~ {\Th}_C+{\cal S}_A~ {\Th}_S)e^{\cal F}.~
\nn\label{solwz1}
\eea
where the Wess-Zumino action has been determined in \bref{solwz}
and $[~~]_{p+1}$ means {p+1}-form coefficient.

Structures and algebras of p+1 bosonic constraints are determined 
only by the Dirac-Born-Infeld part and do not depend on the
form of the Wess-Zumino action;
\bea
\left\{
\begin{array}{ccl}
H&\equiv&\frac12~[~\tp^2~+~\tilde E^a~G_{ab}~\tilde E^b~+~
\tension^2~ {\bf G_F}~]~=~0,\\~~
\\
T_a&\equiv&\tp~\Pi_a~+~\tilde E^b~\CF_{ab}~=~0,~~~~~~~~~~~(a=1,2,...p)
\end{array}\right.\label{defH}
\eea
where  $~p_m,~\z,~E^\mu~$
are canonical momenta conjugate to $X^m$, $\T$ and
$A_\mu$ respectively
and
$~{\bf G_F}=\det (G+\CF)_{ab}.
$
$~~\tilde E^a$ and $\tp_m$ are defined as
\bea
\tp_m&\equiv&p_m~-~\frac{\pa\CL^{WZ}}{\pa \Pi^m_0}~-~
E^a~({\Tb~\gg~\G_m~\pa_a\T})
=\frac{\pa\CL^{DBI}}{\pa \Pi^m_0}
\nn\\
\tilde E^a&\equiv&E^a~-~\frac{\pa\CL^{WZ}}{\pa \CF_{0a}}
=\frac{\pa\CL^{DBI}}{\pa \CF_{0a}}~
\eea
where ${\cal L}$ is regarded as a function of $\Pi_{\mu}$, 
${\cal F}$ and $\theta$.
In addition there appear bosonic U(1) constraints, 
$
E^0~=~0,$ and $~\partial_aE^a~=~0.
\label{gauss}
$
\medskip
The fermionic constraint follows from the definition of 
the momentum $\z$, 
\bea
F~&\equiv&\z~+~p_\l~(\Tb~\G^\l)~\nn\\
&&+~E^a~[~\Tb~\Gamma_{11}\G_\l~
(~\Pi_{a}^\l~ +~ \frac12~\Tb\G^\l \pa_{a}\T~)~
-~ \frac12~\Tb\G^\l(\Tb~\Gamma_{11}\G_\l~\pa_{a}\T)]~-~
(\frac{\pa^r\CL^{WZ}}{\pa \dot \T})~\nn\\
&=&~0.\nn\\&&
\label{defF}
\eea
Both structure and the algebra of the fermionic constraints are
governed by the Wess-Zumino action.

The Poisson bracket of $F$'s is calculated as
\bea
&&\{F_{\alpha}(\sigma),~F_{\beta}(\sigma')\}~=~\int d^p\s~
[~2~(\CC\Xi)_{\alpha\beta}~+~
(\pa_aE^a)~(\Tb\G )_{(\alpha}\cdot(\Tb\gg\G )_{\beta)}]
\delta^p(\sigma-\sigma').\nn\\&&
\eea
Here $\CC$ is the charge conjugation matrix and the symmetric bracket
is defined as $A_{(\alpha \beta)}=
\frac{1}{2}(A_{\alpha \beta}+A_{\beta\alpha})$.
The second term is the Gauss law constraint and $\Xi$ is given
by 
\bea
\Xi&\equiv&\sltp~+~\Gamma_{11}~\slPi_a~\t E^a~~
+~\tension~[{\cal C}_A~e^{\cal F}]_{\bf p}.
\label{defxi}
\eea
Here $[....]_{\bf p}$ is a spatial p form coefficient 
( coefficient of $d\s^1...d\s^p$ ) of the expression $[....]$.
Unlike IIB case the matrix $~\Xi~$ in \bref{defxi} is not nilpotent,
instead there is a zero eigenmatrix;
\bea
\tilde{\Xi}&\equiv&\sltp~-~\Gamma_{11}~\slPi_a~\t E^a~~
-~\tension~[\tilde{\cal C}_{A}~e^{\cal F}]_{\bf p},
\label{deftxi}
\eea
satisfying
\bea
\Xi ~\tilde{\Xi}~=~2H~+~2\hat{\tau}^aT_a~\approx ~0\quad
\label{XiXi0}
\eea
with
\bea
\hat{\tau}^a~=~\t E^a~\Gamma_{11}~
-~{\tension}~\Gamma_{11}~[d\s^a~{\cal S}_A~e^{\cal F}]_{\bf p}\quad .
\label{taua}
\eea
From \bref{XiXi0} it follows 
that the rank of $\Xi$ is one half of 32,
 reflecting the fact that a
half of the fermionic constraints are the first class and the remaining 
half are second class. 

The first class constraint set for IIA D-p-branes is obtained,
using with the zero eigenmatrix $\tilde{\Xi}$ in \bref{deftxi},
\bea
\left\{
\begin{array}{ccl}
\t H&=&H~+~F~\hat{\tau}^a~\pa_a\T,
\\
\\
\t T_a&=&T_a~+~F~\pa_a\T~=~p~\pa_ax~+~\z~\pa_a\T~+~E^b~F_{ab},
\\
\\
\t F&=&F~\tilde{\Xi}\quad.
\end{array}\right.\label{frstcls} 
\eea
$\t H$ and $\t T_a$ generate $(1+p)$ dimensional diffeomorphism.
$\tilde{F}$ is the generator of the kappa symmetry,
\bea
\{ \tilde{F}_{\alpha}(\sigma) ,\tilde{F}_{\beta}(\sigma')\}& =&
-4(C\tilde{\Xi}L)_{\alpha\beta}
\delta (\sigma -\sigma ')~+~\cdot\cdot\cdot
~~,~~
L_{\alpha\beta} \equiv \delta_{\alpha\beta}\tilde{H}
+\hat{\tau}^a_{\alpha\beta}~\tilde{T}_a\quad,
\label{tftfl}
\eea
where $\cdot\cdot\cdot$ contains $\tilde{F}$ and Gauss law constraint.
In case of the Green-Schwarz superstring the constraints $(\tilde{F}_1,
L_{11})$ commute with  $(\tilde{F}_2,L_{22})$, so that right and left moving modes are independent. The D-p-branes are not the case. 

\medskip 

We can perform the irreducible and covariant 
separation of the fermionic constraints 
into the first class and the second class
as was done for the D-string case\cite{MK}:
\bea
&&
\left\{\begin{array}{lccccl}
{\rm  first\ class\ constraints}&:& 
\tilde{F}_1&\equiv&F\tilde{\Xi}\frac{1+\Gamma_{11}}{2}~=~0\\
&&&&&\\
{\rm  second\ class\ constraints}&:& 
{F}_2&\equiv&F\frac{1-\Gamma_{11}}{2}~=~0
\end{array}
\right.\label{sep}\quad.
\eea
The covariant first class constraints in \bref{sep}
make the covariant gauge fixing possible; for example
 one of the chirality
components of $\T$ to be zero \cite{Shgf,RK}. 
By using with the second class constraints $F_2=0$ in \bref{sep},
the Dirac bracket is defined as
\bea
\{A,B\}_D=\{A,B\}-\{A,F_2\}\frac{\Xi_{22}\CC^{-1}}{2T^2{\bf G_F}}\{F_2,B\}~~,~~
\Xi_{22}=\frac{1+\Gamma_{11}}{2}\Xi\frac{1-\Gamma_{11}}{2}\quad.
\eea
The Dirac bracket, or equivalently the separation \bref{sep},
it well defined if
\bea
({\bf G_{F}})~=~\det (G+{\cal F})_{ab}~\neq~0\quad.
\label{gfneq}
\eea
This is the condition for the covariant quantization discussed in \cite{RK}.
If a conformally flat metric can be assumed, $-G_{00}=G_{11}=\cdot\cdot\cdot=G_{pp}$,
${\bf G_{F}}$ is written as
\bea
{\bf G_F}&=&(G_{11})^p+({\cal F}_{ab})^2(G_{11})^{p-2}
+\cdot\cdot\cdot
+({\cal F}_{[a_1a_2}\cdot\cdot\cdot{\cal F}_{a_{p-1}a_p]})^2\quad .
\eea
The necessary condition of the separation \bref{gfneq} is
satisfied by following cases;
\bea
\left\{\begin{array}{lc}
{\rm (i)}&G_{11}> 0\\
{\rm (ii)}&
G_{11}= 0~,~({\cal F}_{[a_1a_2}\cdot\cdot\cdot{\cal F}_{a_{p-1}a_p]})^2>0
\end{array}\right.
~~\Rightarrow ~~
{\bf G_{F}}~>~0\quad.\label{fab}
\eea
(i) are cases in which the static gauge can be taken.
The case (ii) would corresponds to massless particle solutions,
in which the non-zero worldvolume magnetic field,
$({\cal F}_{[a_1a_2}\cdot\cdot\cdot{\cal F}_{a_{p-1}a_p]})^2 >0$,   
guarantees the separation \bref{sep}.

There is 
alternative necessary condition;
the condition of the DBI action to be well defined,
$-\det (G+{\cal F})_{\mu\nu}\geq 0$.
For D-string case this necessary condition adding to 
\bref{gfneq} reduces to that:
if ${E}^{1}\neq 0$ then $G_{11}\neq 0$ \cite{MK}. 
In other words, ${E}^{1}\neq 0$ guarantees both 
the separation and the static picture of a D-string.
However for IIA D-p-branes for p$>1$
the situation is different from the D-string,
the massless solutions can not be excluded only by 
the condition of the DBI action $-\det (G+{\cal F})_{\mu\nu}\geq 0$. 
\vskip 6mm
 

\section{Global supersymmetry (SUSY)}
\indent
\medskip
The global supersymmetry transformations of $X$, $\T$ and $A$ are
\bea
&&\D_\ep~\T=\ep,~~~~~~~~\D_\ep~X^m~=~\ba\ep~\G^m~\T  \nonumber\\
&&
\D_\ep~A=(\ba\ep~\Gamma_{11}~\G~\T)_m~d~X^m~-~\frac16~\ba\ep~
\Gamma_{11}~V\cdot \Gamma~\T\quad ,
\eea
so that $d\theta$, $\Pi$ and 
$\CF$ are SUSY invariant.
The canonical supersymmetry generator is
\bea
Q~\ep&=&\int d\s^p~(p_m~\D_\ep~X^m~+~\z~\D_\ep~\T~+~E^a~\D_\ep~A_a)~-~
\int d\s^p~U^0_\ep.
\label{qeqe}
\eea
$U^0_\ep$ is determined from the surface term of the SUSY variation of 
the Wess-Zumino lagrangian,
\bea
\D_\ep~L^{WZ}~\equiv~d~(~U_\ep~)~~~~~,~~~~~U^0_\ep~=~[~U_\ep~]_{\bf p}\quad.
\label{u0u0}
\eea

Under the SUSY transformation the RR potential transforms as 
\bea
\D_\ep~C~=~(d~D_\ep)~e^B
\eea
where $D_\ep$ is even form for IIA.
The Wess-Zumino action transforms by an exact form with
\bea
U_\ep~=~D_\ep~e^{dA}.\label{qwzDF}
\eea
Explicit expression of ${ D}_\ep$ is calculated as  
\bea
{ D}_\ep&=&\{
\bar{\hat{\Theta}}_C~(~{\cal C}_{A}~\delta_{\epsilon}\hat{\Th}_C
 		     +~{\cal S}_{A}~\delta_{\epsilon}\hat{\Th}_S)
+\bar{\hat{\Theta}}_S~(-~\tilde{{\cal S}}_{A}~\delta_{\epsilon}\hat{\Th}_C
 		     +~\Gamma_{11}~\tilde{{\cal C}}_{A}~
\delta_{\epsilon}\hat{\Th}_S)
\}{\rm e}^{B}. \nn\\&&
\label{solcalD}
\eea
The conserved SUSY charges are obtained completely from 
\bref{qeqe}, \bref{u0u0}, \bref{qwzDF}, \bref{solcalD}.

\medskip

The poisson bracket of \bref{qeqe} is calculated as
\bea
\{~Q_\ep,~Q_{\ep'}~\}&=&\int d\s^p~[~2~(\ba\ep~\G~\ep')~p~+~
2~(\ba\ep~\Gamma_{11}~\G~\ep')~\pa_a X~E^a
\nn\\
&-&\frac12(\pa_a~E^a)~\{(\Tb\Gamma_{11}\G\ep)\cdot(\Tb\G\ep')~-~
(\Tb\Gamma_{11}\G\ep')\cdot(\Tb\G\ep)\}~]
\nn\\
&-&2\tension~\int d\s^p~\bar{\epsilon}[{\cal C}_A(\sldX)
e^{dA}]_{\bf p}\epsilon'\quad .
\label{QeQe}
\eea
In the last term we left only contribution which could remain 
for the non-trivial topological configuration of $X$ and $A$.

In 10-dimensions there are $10+1+10+45+210+252=\frac{32\times 33}{2}=528$ 
independent symmetric Gamma matrices;
\bea
&&
\CC\Gamma_M=
\{~\CC\Gamma_m,~\CC\Gamma_{11},~\CC\Gamma_m\Gamma_{11},~
\CC\Gamma_{m_1m_2},~
\CC\Gamma_{m_1m_2m_3m_4}\Gamma_{11},~\CC\Gamma_{m_1m_2m_3m_4m_5}~\} ,
\nn\\&&
\eea
Using this basis \bref{QeQe} are decomposed to define
the central charges,
\bea
&&\{~Q_{\A },~Q_{\B }~\}\nn\\
&&~~~~~=
-~2~(\CC\Gamma_m)_{\A\B}~{\cal P}^m
\nn\\
&&~~~~~~~-2~(\CC\Gamma_{11}\Gamma_m)_{\A\B}~\int d\s^p~(~\pa_a X^m~E^a~)~
\nn\\
&&~~~~~~~+\int d\s^p(\pa_aE^a)~[
-\frac{10}{32}~(\CC\Gamma_{11})_{\A\B}~
(\Tb\T)
+\frac{6}{32}~(\CC\Gamma_{m_1m_2})_{\A\B}~
(\Tb\Gamma_{11}\Gamma^{m_2m_1}\T)
\nn\\
&&~~~~~~~~~~~~~~~~~~~~~~~~~
-\frac{2}{32}~(\CC\Gamma_{m_1m_2m_3m_4}\Gamma_{11})_{\A\B}~
(\Tb\Gamma^{m_4m_3m_2m_1}\T) ]
\nn\\
&&~~~~~~~-2\tensionm(\CC\Gamma_{M})_{\A\B} Z^M\quad,
\label{qqal}
\eea
with
\bea
Z^M=\frac{1}{32}~
{\rm tr}~\int~\left(\Gamma^M ~[{\cal C}_A(\sldX)e^{dA}]_{\bf p}\right)
\quad.
\eea
Here $\Gamma^M$ is inverse of $\Gamma_M$ ,
${\rm tr}(\Gamma^N{\CC}^{-1}\CC\Gamma_M)=32\delta^N_M$ and  
${\cal P}^m$ in the first term is the total momentum.
The SUSY central charges contain membrane charges
and 
topological charges for the worldvolume gauge field.
The possible interpretation of the central charges have been given in
\cite{Hammer}. 
We discuss it further in the next section.
The SUSY central charges, $Z^M$,  are listed in the table 2.
The vector indices of $Z^M$ are abbreviated,
for examples $Z^{[2]}$ in place of $Z^{m_1m_2}$.

\begin{table}[hbtp]
\caption[SUSYcc]{SUSY central charges: 
$Z^{M}$}

\label{tblz}
\begin{center}
\begin{tabular}{|c|c|c|c|}
\hline
 & $Z^{11}$ & $Z^{[2]}$ & $Z^{[4]11}$ \\
\hline
D-0&1&$0$&$0$\\
\hline
D-2&$dA$&$(dX)^2$&$0$\\
\hline
D-4&$(dA)^2/2!$&
$(dX)^2~dA$&
$(dX)^4$\\
\hline
D-6&$(dA)^3/3!$&
$(dX)^2~(dA)^2/2$&
$(dX)^4~(dA)+^*(dX)^6$\\
\hline
D-8
&$(dA)^4/4!$
&$(dX)^2~(dA)^3/3!+^*(dX)^8$&
$(dX)^4~(dA)^2/2+^*(dX)^6~dA$\\
\hline
\end{tabular}
\end{center}
\end{table}

In the rest frame 
IIA D-p-branes the SUSY algebra 
\bref{qqal} is written as 
\bea
\{Q_{\alpha},Q_{\beta}\}=-2
\left((\CC\slcP)_{\alpha'\beta'}
\otimes {\bf 1}_{AB}+
\sum_{i=1}^3(a_i)_{\alpha'\beta'}\otimes (\tau_i)_{AB}
\right)
\label{QQp0bps}\nn\\
\eea
with 
\bea
\left\{ \begin{array}{ccl}
a_1&=&\tensionm\CC\gamma_{[2]}Z^{[2]}\\
a_2&=&\tensionm i\CC Z^{11}+\tensionm i\CC\gamma_{[4]}Z^{[4]11}\\
a_3&=&\CC\gamma_m \int \partial_a X^m E^a
\end{array}\right.\quad.
\eea
for chiral gamma matrices $\gamma_m$. ${\bf 1}$ and $\tau_i$ are $2\times 2$ 
matrices whose indices represent chiralities.
The above SUSY algebra \bref{QQp0bps} can be diagonalized because of 
 symmetric indices.
In order to examine the bound of the SUSY algebra,
 we assume $[a_3,a_1+ia_2]=0$
and use the fact of ${\rm tr}[a_1,a_2]=0$,
and also assume that
the second part of the right hand side of
\bref{QQp0bps} can be diagonalized as $\lambda\tau_3$.
The assumption $[a_3,a_1+ia_2]=0$
reduces to
$E^a=0$.
For a situation 
with $\partial_{\mu}X^m=\delta_{\mu}^m$,
$\theta=0$
and constant U(1) gauge fields, 
the SUSY algebra \bref{QQp0bps} becomes in
the rest frame with $\CC=i\Gamma_0$ 
\bea
({\cal P}_0  {\bf 1}+\lambda\tau_3)~|~\rangle \geq  0\quad,
\quad \lambda^2 =-\frac{1}{16}{\rm tr}~\sum_i(a_i)^2
=V_{(p)}^2~\tensionm^2{\bf G_F}
\label{bps2a}
\eea
analogous to the D-string case \cite{MK}. 
$V_{(p)}$ is the p-brane volume.
By using the $\tau$ diffeomorphism constraint 
of the D-p-brane \bref{frstcls},
${\cal P}_0=\lambda$ is shown under the same assumption,
\bea
\int d^p\sigma \tilde{H}=-\frac{1}{2}\frac{1}{V_{(p)}}
\left[-({\cal P}_0)^2+\lambda^2\right]=0
\quad.\label{bpstdf}
\eea
This confirms the BPS saturated state \bref{bps2a}.
It suggests that groundstates of IIA D-p-branes, 
which are BPS saturated states,
satisfy the above assumptions;
$\partial_{\mu}X^m=\delta_{\mu}^m$,
$\theta=0$,
 $F_{ab}=$const. and $E^a=0$.
In the BPS saturated state, \bref{bps2a} becomes
the projection operator $\lambda({\bf 1}+\tau_3)$,
then it 
leads to that only N=1 SUSY survived.

\vskip 6mm
\section{ Summary and Discussions}\par
\indent

In this paper we obtained the Wess-Zumino
action for IIA D-p-branes in an explicit form. 
A compact expression is 
obtained by choosing the spinor variables 
\bref{defThCS} analogous to IIB case \cite{KH}.
The fermionic constraints can be separated into the first class and the 
second class in a covariant and irreducible way.
The first class constraints generate the local supersymmetry (kappa symmetry).
In contrast with
 the case of the fundamental superstring,
D-p-brane picture has massive ground states and allow the static gauge.
This assumption of the D-brane leads to
the covariant and irreducible kappa generator. 
We have derived the necessary condition of the well-defined separation in 
\bref{gfneq}, ${\bf G_F}\neq 0$ \cite{RK}.
 For the super-D-string we have shown that $({\cal F}_{01})^2>0$
guarantees both the static string ground state and 
the well-defined separation:
 \cite{MK}.
This quantity ${\cal F}_{01}$ 
($E^1$ in canonical variable)
is interpreted as non-trivial winding of the ${\bf S}^1$ \cite{Wbsp}. 
In other words, the winding of ${\bf S}^1$ 
makes groundstates to be massive and string-like, and
also makes the well-defined separation possible.  
On the other hand for IIA D-branes the role of $E^a$ may be different from the one for the D-string.

The global supersymmetry (SUSY) charges and SUSY algebra are calculated
for general IIA case. 
The SUSY algebra contains topological charges with a typical form
\bea
Z^{[q]}~\sim~\int ~(dX)^{q}~(dA)^
{\frac{p-q}{2}},~~~~~~~~(q=0,2,...,p).\label{tpch}
\eea
The $q=p$ term is a membrane charge and measures
 the topologically non-trivial worldvolume.
$q<p$ charges correspond to the topological charges for 
the gauge field.
For $p=2,~q=0$ \bref{tpch} allows Dirac monopole 
configuration in $2+1$ dimension \cite{Hammer}. 
For $q=0$ \bref{tpch} allows monopole configuration in $p+1$ dimensions 
if the gauge group $G$, which gives nontrivial homotopy 
class $\Pi_p(G)$ \cite{Nkhr}, is induced by the Chan-Paton factor 
\cite{pol,Wsdv}. 
It is interesting to consider the case where the worldvolume gauge fields 
get masses by some mechanism \cite{pol,Doug}.
In such cases for $q=p-2$, \bref{tpch} represents
 a $q$-dimensional defect for D-p brane.
For example,
 a surface defect for D-4 brane,
 a 4-dimensional defect for D-6 brane,
 and so on.
It will also represents 
$q$-dimensional defects if non-Abelian  
$A\in G$ is induced where $G$ belongs to non-trivial homotopy class
${\Pi}_{p-q-1}(G)$. 
These defects may be interpreted as solitonic configurations 
of brane coordinates arising from intersections \cite{HW,CHS,Doug}. 

In this paper we have considered D-p-branes only in the trivial
background.
In case of type IIA there exists an alternative background, 
namely massive IIA supergravity \cite{Rmn}.
The  massive IIA supergravity background allows
the Chern-Simon terms \cite{BRoo,GHT,BCT}. 
The mass parameter is 
interpreted as the square root of the cosmological constant \cite{BRGPT}. 
The coefficient of the Chern-Simon term is quantized as usual, 
and so is the cosmological constant. 
The quantization of the cosmological 
constant is consistent with the duality argument \cite{GHT}. 
The RR gauge-invariant field strength for this case 
has the following mass dependence \cite{BRGPT}, 
\bea
dL^{WZ+CS}~=~\tensionm~R~e^{\cal F}~~~,~~~ 
R~=~dC-HC+me^{B}~
\label{wzcs}
\eea
The mass dependent term in \bref{wzcs} gives the Chern-Simon term. 
Although $R$ has mass dependence, it again satisfies the same 
Bianchi identity as the one without mass term \bref{Bianchi}.  
The previous works related to the Chern-Simon term have been 
done in the superspace language.
It is interesting to examine how the Chern-Simon term is 
incorporated with the kappa symmetry 
and supergravity in the component language.
We leave this issue for future investigation.

\vskip 6mm
{\bf Acknowledgements}\par
\medskip\par
 M.H. thanks for 
helpful conversation for K.Hashimoto and N.Ishibashi.
M.H. is partially supported by the Sasakawa Scientific Research Grant 
from the Japan Science Society.

\end{document}